\documentclass[runningheads]{llncs}
\usepackage[T1]{fontenc}
\usepackage{amsmath,amsfonts}
\usepackage{xcolor}
\usepackage{algpseudocode}
\usepackage{algorithm}
\usepackage{caption}
\usepackage{graphicx}
\usepackage{color}

\newcommand{\remove}[1]{}
\usepackage{graphicx}
\usepackage{textcomp}
\usepackage{hyperref}
\usepackage{xcolor}
\begin{document}
\title{Enhancing MOTION2NX for Efficient, Scalable and Secure Image Inference using Convolutional Neural Networks}
%
%

\author{Haritha K \inst{1}\and
Ramya Burra \inst{2} \and
Srishti Mittal\inst{1} \and 
Sarthak Sharma\inst{1} \and
Abhilash Venkatesh\inst{1} \and
Anshoo Tandon\inst{1} }
\authorrunning{Haritha K et al.}
%
\institute{IUDX, FSID, IISc Bengaluru \\
\email{\{gattu.haritha,  srishti.mittal413, sarthaksharma070, abhilashnitk3, anshoo.tandon\}@gmail.com}\\
 \and
CBR, IISc Bengaluru\\
\email{ramya.burra@gmail.com}
}
\maketitle             
\vspace{-1cm}
\begin{abstract}
This work contributes towards the development of an efficient and scalable open-source Secure Multi-Party Computation (SMPC) protocol on machines with moderate computational resources. We use the ABY2.0 SMPC protocol implemented on the C++ based MOTION2NX framework for secure convolutional neural network (CNN) inference application with semi-honest security. Our list of contributions are as follows. Firstly, we enhance MOTION2NX by providing a tensorized version of several primitive functions including the Hadamard product, indicator function and argmax function. Secondly, we adapt an existing Helper node algorithm, working in tandem with the ABY2.0 protocol, for efficient convolution computation to reduce execution time and RAM usage.
Thirdly, we also present a novel splitting algorithm that divides the computations at each CNN layer into multiple configurable chunks. This novel splitting algorithm, providing significant reduction in RAM usage, is of independent interest and is applicable to general SMPC protocols. 
\vspace{-0.25cm}
\vspace{-0.3cm}
\end{abstract}
\vspace{-0.3cm}
\section{Introduction}
\vspace{-0.3cm}
    In today’s data-driven world, ensuring privacy and security in computation is essential, as privacy is a fundamental human right~\cite{PrivFundamental}. Secure Multi-Party Computation (SMPC) addresses these concerns by enabling secure data sharing across various domains. This paper focuses on optimizing SMPC for real-world scenarios, particularly on low-resource machines, to reduce execution time and memory use. Our convolutional neural network (CNN) inference solution on the MNIST dataset, with semi-honest security~\cite{semi-honest}, runs on virtual cloud machines with less than $0.5$~GB RAM, resulting in substantial cost reduction~\cite{amazon-pricing}.

To achieve resource optimization, we modified the C++ based MOTION2NX framework~\cite{braun2021a}, adding new functionalities for secure inference tasks. We addressed limitations such as memory inefficiencies, and our enhancements include more efficient tensor operations, novel approaches to handle missing functions like argmax, and using a third-party Helper node to reduce memory usage and execution time. These improvements significantly enhance the framework’s efficiency while ensuring data privacy and integrity (see Section~\ref{section : Contributions to MOTION2NX}). Further, we made our implementation fully scalable in terms of RAM usage by splitting computations at the convolution and neural network layers. This approach ensures that machine learning models can execute efficiently within the available memory resources.

We use the data provider framework of SMPC expalined in~\cite{Burra}. We consider two compute servers for executing the ABY2.0 SMPC protocol~\cite{ABY2.0} and two data providers that possess input data and machine learning model, respectively (see Section~\ref{dataprovider model}). We assume that the readers are familiar with ABY2.0 protocol~\cite{ABY2.0}.
\vspace{-0.55cm}
\subsection{Related Work}
\vspace{-0.25cm}
Over the past several years, there has been an increased focus on practical application of SMPC to real-world problems e.g., secure auction, secure gender wage gap study. In secure auction, to preserve bid privacy, a three-party SMPC system was employed~\cite{smpc-live}. 
Secure gender wage gap study uses SMPC techniques to ensure collective computation of aggregate compensation data while preserving individual privacy~\cite{gender-wage}. This approach empowers organizations to collaborate effectively while upholding data privacy. 

We remark that the above two described SMPC applications are not memory or computationally intensive. We present modifications and functional additions to the MOTION2NX framework for secure neural network inference, addressing memory and computational intensive tasks. These updates aim to support secure disease prediction (see~\cite{secure-nn} for secure medical image analysis), where one party shares medical images and the other shares a pre-trained neural network model.
\vspace{-0.50cm}
\subsection{Our Contributions}
\vspace{-0.20cm}
The following is the list of our contribution, specifically to MOTION2NX setup.
\vspace{-0.55cm}
\begin{itemize}
\vspace{-0.05cm}
\item We enhance MOTION2NX by providing a tensorized version of several primitive functions including secure Hadamard product, secure indicator function and secure argmax function. Our design of the secure indicator function is based on a novel approach that uses SecureRelu function available in the baseline MOTION2NX implementation. The secure indicator function is used, in turn, as a building block for a novel implementation of the secure argmax function.  Further, secure indicator function can be used to design secure piece-wise linear approximation of a function.
\item We develop a novel splitting algorithm that divides the computations at each CNN layer into multiple configurable chunks. This algorithm provides significant reduction in RAM usage and is applicable to general SMPC protocols. 
\item We adapt an existing Helper node algorithm, working in tandem with the ABY2.0 protocol, for efficient convolution computation. This algorithm not only reduces execution time but also reduces the RAM usage required to execute CNN models, but comes at a cost of an additional compute server. 
\item To simplify the end-to-end image inference application on the cloud machines, we develop the docker images which contain all the essential dependencies and compiled binaries.

\end{itemize}
\vspace{-0.1cm}
\color{black}
\vspace{-0.3cm}
\section{Preliminaries}
\vspace{-0.2cm}
In this section, we provide preliminary details of our secure convolutional neural network inferencing implementation. We consider MOTION2NX, a C++ framework for generic mixed-protocol secure two-party computation in our paper.
We discuss the details of our enhancements and optimizations in Section~\ref{section : Contributions to MOTION2NX}.
\vspace{-0.2cm}
\subsection{$N$-layer Convolutional Neural Network}\label{subsec:CNN}
\vspace{-0.2cm}
Our optimizations in MOTION2NX enable efficient execution of deep CNN. For illustrative purposes, we present the details for $4$-layer CNN on MNIST. A similar procedure can be adapted for any other dataset, using a corresponding pre-trained CNN model with arbitrary number of network layers.

In MNIST inference task, the input to the neural network is a real-valued vector. The output $i$ is an integer, $i \in \{0 \dots 9\}$. For illustrative purposes, we consider $4$-layer model with two convolution layers and two fully connected layers for MNIST dataset inferring task with parameters as described in Table~\ref{Table:Two_Layer_CNN_parameters}.
\vspace{-0.5cm}
\begin{table}
\vspace{-0.15cm}
\caption{Neural Network Configuration used for Inferencing MNIST Data}
\centering
\begin{tabular}{|c|c|c|c|c|c|} 
\hline      
{Layer } & {No. of Kernels}& Padding &Strides& {No. of parameters } & {Biases} \\ 
\hline
{CNN1} & 5 & [1, 0, 1, 0] & [2,2]&{5 $\times$ 1 $\times$ 5 $\times $5} & {5 $\times$ 1}   \\ 

\hline
{CNN2} & 3&[1, 0,1,0]&[1,1] &{3 $\times$ 5 $\times$ 4 $\times $4} & {3 $\times$ 1}   \\

\hline
{NN1} & {100 neurons} &-&-&{108 $\times$ 100} &{100$\times 1$}\\
\hline
{NN2} & {10 neurons}  &-&-&{100 $\times$ 10} &{10$\times 1$}\\
\hline
\end{tabular}
\label{Table:Two_Layer_CNN_parameters}
\vspace{-0.5cm}
\end{table}

Algorithm~\ref{alg:CNN_inferencing} describes a simple two convolution layer and two neural network layer inference implementation with ReLU activation in secure mode. This algorithm takes ABY2.0 shares of input data (image), network weights, and biases as inputs and produces ABY2.0 shares of the predicted label as output. The image shares, weight shares and bias shares at server-$i, i \in \{0, 1\}$ are represented as $x^i, w^i_j$ and $b^i_j, j \in \{1, 2, 3, 4\}$ respectively. Further, $ w^i_1$ and $ w^i_2$ represent the kernel parameters for CNN layers while $ w^i_3$ and $ w^i_4$ represent matrix parameters for fully connected neural network layers.

We recall that ABY2.0 shares consist of a pair comprising a public share and a private share~\cite{ABY2.0}. For instance, the ABY2.0 shares of an input variable $y$ associated with server $i$ are represented as a pair consisting of $\Delta_y$ and $[\delta_{y}]_i$. Note that $\Delta$ and $\delta$ represent public share and private share respectively. We remark that the steps outlined in Algorithm~\ref{alg:CNN_inferencing} can be readily extended to secure execution of general deep CNN inferencing tasks (with any number of layers). Most of the secure functions listed in Algorithm~\ref{alg:CNN_inferencing} were provided by MOTION2NX framework. We used ABY2.0 arithmetic protocol for convolution, multiplication and addition. We used Yao protocol to perform the ReLU function, as Yao performs better for comparison operations. 
Algorithm~\ref{alg:CNN_inferencing}, unfortunately, couldn't be executed completely in MOTION2NX using its baseline built-in functions as SecureArgmax was not available. We enhance the MOTION2NX framework by implementing a tensorized version of the SecureArgmax function as explained in Algorithm~\ref{alg:tensor-argmax} in Section~\ref{section : Contributions to MOTION2NX}. This marks a departure from the previous approach~\cite{Burra}, where the authors relied on a non-tensor version of SecureArgmax to achieve a similar functionality. 
\vspace{-0.4cm}
\begin{algorithm}

    \caption{CNN inferencing task with ReLU activation function at compute server-$i$, $i \in \{0,1\}$}
    \label{alg:CNN_inferencing}
    \begin{algorithmic}[1]
        \Require Image shares $x^i$, CNN weight shares $w^i_1, w^i_2$, fully connected NN weight shares $w^i_3, w^i_4$, bias shares $b^i_1, b^i_2, b^i_3, b^i_4$. All the inputs are in the form of ABY2.0 share vectors.
        \Ensure Shares of predicted class label $\hat{y}^i$
        \For{$j \gets 1$ to $2$}  \\
        \quad $z^i_j = $SecureCNN($w^i_j,x^i,b^i_j)$ \\
        \quad $h^i_j = $SecureReLU$( z^i_j)$ 
    \EndFor
       \For{$j \gets 3$ to $4$}\\
       \quad $z^i_j = $SecureAdd(\,SecureMul$(w^i_j,h^i_{j-1}), \,b^i_j)$ \\
         \quad $h^i_j = $SecureReLU$( z^i_j)$
       \EndFor
        \State Compute predicted class label: \\
        \quad $\hat{y}^i= $SecureArgmax($ h^i_4$) 
        \State \textbf{Return} $\hat{y}^i$
        \end{algorithmic}
        \vspace{-0.1cm}
\end{algorithm}

\subsection{Data Provider Model}
\label{dataprovider model}
\vspace{-0.2cm}
We adopt the Data Provider model from~\cite{Burra}. In this model the data providers (Image provider and Model provider) create shares of their private data and send them to the compute servers. Compute servers perform the inference task and send the output shares to the Image provider. Compute servers are unaware of the clear output result. For secure inference task, we consider that the neural network model is pre-trained and is proprietary to a model provider. Similarly, the image for the inference task is private to image data provider.
 \vspace{-0.2cm}
\section{Enhancements to MOTION2NX Framework} \label{section : Contributions to MOTION2NX}
\vspace{-0.2cm}
In this section, we discuss our enhancements to the MOTION2NX framework. The baseline framework unfortunately did not provide certain functions necessary for executing the inference task, e.g., argmax. In this work, we designed an efficient argmax implementation that, in turn, is based on the use of the Hadamard product and indicator function. We present these implementations below.

\vspace{-0.3cm}
\subsection{Hadamard Product:} 
\begin{algorithm}
   \caption{Protocol SecureHadamard$(⟨a⟩, <b>)$}
   \label{alg:Hamm}
   \begin{algorithmic}[1]
   
  \State \textbf{Setup Phase:}   
    \State Each party $P_i, i \in \{0, 1\}$ compute the following
    \State $[\Delta_y]_i = [\delta_a]_i \odot [\delta_b]_i$
    \State $[\Delta_y]_i = [\Delta_y]_i + \textbf{OT}\Big([\delta_a]_i\odot[\delta_b]_{1-i}\Big)+\textbf{OT}\Big([\delta_a]_{1-i}\odot[\delta_b]_i\Big)$
    
  \State \textbf{Online Phase:} 
    \State Each party $P_i, i \in \{0, 1\}$ compute the following
    \State $[\Delta_y]_i = [\Delta_y]_i - [\Delta_a] \odot [\delta_b]_i - [\Delta_b] \odot [\delta_a]_i$
    \State $[\Delta_y]_i = [\Delta_y]_i + i*\Big([\Delta_a] \odot [\Delta_b]\Big) $
    \State perform truncation operation on $[\Delta_y]_i$
    \State $[\Delta_y]_i = [\Delta_y]_i + [\delta_y]_i$, where $[\delta_y]{}_i \in_R \mathbb{Z}_{2^{64}}$ 
    \State $P_i$ sends $[\Delta_y]_i$ to $P_{1-i}$
    \State Both $P_0$ and $P_1$ calculate $\Delta_y = [\Delta_y]_0 + [\Delta_y]_1$
   \end{algorithmic}
   \vspace{-0.1cm}
\end{algorithm}
\vspace{-0.2cm}
Let $⟨a⟩$ and $⟨b⟩$ represent the ABY2.0 shares of matrices $a$ and $b$ respectively. Let $a \odot b$ denote the Hadamard product of $a$ and $b$. We implemented Hadamard product on these matrices as explained in Algorithm~\ref{alg:Hamm}. Note that, \textbf{OT} block represents oblivious transfer that is required for secure multiplication~\cite{ABY2.0}.
Implementing the \textbf{OT} block for the Hadamard product is not straightforward. To achieve this, we first studied how \textbf{OT} is implemented for matrix multiplication in the existing MOTION2NX framework. Then, we made the necessary modifications to adapt it for the Hadamard product.

Further, we extended this implementation to accommodate scenarios where matrix $a$ is in clear form while matrix $b$ is in ABY2.0 shares. 

\vspace{-0.2cm}
\subsection{Indicator Function: } 
\vspace{-0.15cm}
Let $I()$ denote indicator function in clear. We implemented its secure version by leveraging the SecureRelu(.) function provided in the baseline MOTION2NX framework, as detailed below. The indicator function in clear $I()$ is given by,
\vspace{-0.10cm}
\begin{equation}\label{eq:IndClear}
    I(x) = \begin{cases}
        0 & \text{if } x < 0 ,\\
        1 & \text{otherwise.}
    \end{cases}
    \vspace{-0.1cm}
\end{equation}

\begin{figure}
    \centering
    \vspace{-0.55cm}
    \includegraphics[width=4cm,height=3cm]{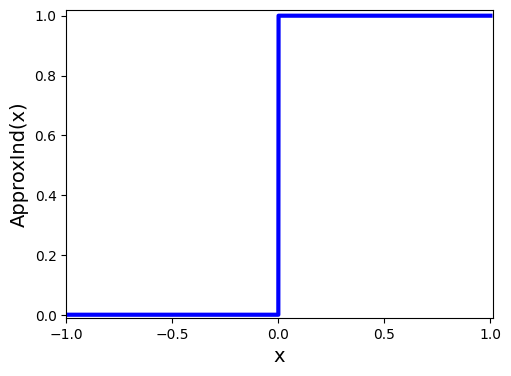}
\vspace{-0.3cm}
    \caption{ApproxInd(x) for $K = 2^{14}$}
    \label{fig:ApproxInd}
    \vspace{-0.55cm}
\end{figure}

We use the Relu function as a building block to implement the indicator function \eqref{eq:IndClear}. This is achieved by constructing an approximate indicator function as follows,
\vspace{-0.10cm}
\begin{equation} \label{eq:ApproxInd}
    \text{ApproxInd}(x) = \text{Relu}(1-K\text{ Relu}(-x))
\end{equation}
where $K > 2^{13}$. We plot ApproxInd$(x)$ for $K = 2^{14}$ in Fig.\ref{fig:ApproxInd} for $-1 \leq x \leq 1$. Note that, ApproxInd$(x)$ can also be written as
\vspace{-0.0cm}
\begin{equation} \label{eq:2_approxind}
    \text{ApproxInd}(x) = \begin{cases}
        0, & \text{if } x<-\frac{1}{K} ,\\
        1+ Kx, & \text{if }-\frac{1}{K} \leq x \leq 0,\\
        1 & \text{if } x > 0.    
    \end{cases}
\end{equation}
Further, MOTION2NX uses the ring $\mathbb{Z}^{2^{64}}$ for arithmetic operations. It supports $64$ bit fixed point arithmetic including $f$ bits to represent the fraction part of variable $x$. In our implementation we chose $f = 13$.
When $x$ is a fixed point number (with $f = 13$ bits for the fractional part) and $K > 2^f = 2 ^{13}$ then comparing \eqref{eq:IndClear} and \eqref{eq:2_approxind}, we observe that $I(x) = \text{ApproxInd}(x)$. This means that our ApproxInd$(x)$ does not result in any approximation error. Using these ideas we implement the secure version of the Indicator function as follows.
\vspace{-0.3cm}
\begin{equation}\label{eq:secure_indicator}
    \text{SecureInd}(x) = \text{SecureRelu}\big(1-K\text{ SecureRelu}(-x)\big),
    \vspace{-0.3cm}
\end{equation}
\vspace{-0.1cm}
where $K>2^{f}$, $f $ is number of fractional bits. 
\vspace{-0.1cm}
\subsection{Argmax: }
\vspace{-0.2cm}
\par Our novel implementation of the tensor version of the secure argmax function is based on Hadamard product and Indicator function (discussed in the previous subsections), and also uses the SecureMaxPool function from the baseline MOTION2NX framework. We explain the implementation of SecureArgmax in Algorithm~\ref{alg:tensor-argmax}, and compare the execution times of its scalar and tensor versions in Table~\ref{tab: Execution Time Comparision for Scalar and Tensor Versions of Argmax}. The input to SecureArgmax is ABY2.0 shares tensor and output is the shares of the highest index at which the maximum element is present.

In step~$1$, we determine the maximum value in  the input array of length $n$ using existing SecureMaxpool function. In step~$2$, we create an array of length $n$ where each element equal to the maximum value computed in step~$1$. In step~$3$, we subtract this array from the input array, and in step~$4$, we pass the result to an indicator function. At this stage, we have an array with zeros and ones, where the positions of ones indicate the presence of the maximum value in the input array. Moving on to step~$6$, we perform Hadamard multiplication of the array obtained from step~$4$ with $[0,1, \dots, n-1]$. Note that the non-zero elements present in $z_4^i$ are the indices at which the maximum element is present. However, we only need one index, so we select the highest index at which the maximum element is present (by executing SecureMaxpool), as outlined in step~$7$. 
\vspace{-0.1cm}
\begin{algorithm}

    \caption{SecureArgmax: Computation of shares of the index of the maximum element in the input array at compute server-$i, i \in {0, 1}$}
    \label{alg:tensor-argmax}
    \begin{algorithmic}[1]
        \Require ABY2.0 shares of input vector $x^i$ and its length $n$ 
        \Ensure Shares of the highest index of the maximum element, $\hat{y}^i$
        \State Compute the shares of the maximum element $z^i_1$ as 
        \quad $z^i_1 = $ SecureMaxPool$(x^i)$
       \State Let $t$ be all ones vector of size $n$. Create an array $z^i_2$ of size $n$ as
        \quad $z^i_2 = $ SecureConstMatrixMult$( z^i_1, t)$ 
        \State Subtract the maximum value from the input array: 
        \quad $z^i_3 = $ SecureAdd$(x^i,$ SecureNegate$ (z^i_2))$
        \State Compute the array with zeros and ones, where one represents the presence of maximum element at that position: 
         \quad $z^i_4 = $ SecureInd$(z^i_3)$ 
        \State Let $k$ denote the vector $[0,1,2, \dots, n-1]$. Compute the argmax shares by executing the following two steps: \\
        \quad $z^i_5= $ SecureConstHadamard$(k, z^i_4)$\\
        \quad $\hat{y}^i = $ SecureMaxPool$(z^i_5)$
        \State \textbf{Return} $\hat{y}^i$

    \end{algorithmic}
\end{algorithm}
\vspace{-0.0cm}
\par In Table~\ref{tab: Execution Time Comparision for Scalar and Tensor Versions of Argmax}, we provide the execution times for the scalar and vectorized versions of Argmax. To facilitate a comparison between the scalar and tensor versions, we use vectors of varying lengths and measure the execution times. We observe that the tensor version of Argmax exhibits lower execution times in comparison to the scalar version.
\vspace{-0.50cm}
\begin{table}[h]
\caption{Comparison of Argmax Execution Time for Scalar and Tensor Versions}
    \centering
        \begin{tabular}{|c|c|c|}
        \hline
        {Vector Length} & \multicolumn{2}{|c|}{Execution Time (sec)} \\
        \cline{2-3}{} & {Scalar Version} & {Tensor Version} \\
        
        \hline
        100  & 11.23 & 4.46 \\
        \hline
        200  & 22.95 & 5.025 \\
        \hline
        300 &  35.6 & 8 \\
        \hline
        400 &  73.7 & 11.2 \\
        \hline
        500 &  154 & 25.25 \\
        \hline
    \end{tabular}
    
    \label{tab: Execution Time Comparision for Scalar and Tensor Versions of Argmax}
    \vspace{-0.2cm}
\end{table}
\vspace{-0.2cm}
\section{Numerical Results:}
\label{sec:cnn_inference_discussion}
\vspace{-0.3cm}
\par In this section, we illustrate the RAM usage and execution time taken by Algorithm~\ref{alg:CNN_inferencing} on MNIST dataset. Towards this, first we explain the setup for the execution and subsequently the challenges we encountered in image inferencing and the techniques we employed to overcome these challenges.

\par 
In practice, compute servers are usually hosted in the cloud on different LANs. To measure execution time, we deployed compute server 0 on Microsoft Azure and compute server 1 (along with the Helper node) on AWS (see Table~\ref{tab:cloud_config}). The image provider and model/weights provider operated on separate local machines.

\begin{table}[h]
\vspace{-0.7cm}
\caption{Cloud Configuration}
\centering
\begin{tabular}{|c|c|c|}
        \hline

Server 0 &
Azure: b1s 1vcpu, 1 GB RAM, 30 GB SSD \\
\hline
Server 1 & 
AWS: t2.micro 1vcpu, 1 GB RAM, 30 GB SSD \\
\hline
Helper node  & 
AWS: t2.nano 1vcpu, 0.5 GB RAM, 30 GB SSD \\
\hline
Image provider & Personal laptop \\
\hline
Weights provider & Personal laptop \\
\hline
\end{tabular}
\label{tab:cloud_config}
\vspace{-0.5cm}
\end{table}

\par  
The execution of Algorithm~\ref{alg:CNN_inferencing} on MNIST dataset as a single process required more than $1$~GB RAM on cloud machines for successful execution of the CNN inference application. On the other hand, the independent execution of each layer in the network consumed $0.219$~GB RAM (see first row in Table~\ref{Table:Split Two_Layer_CNN MNIST}). Note that, the image size in the MNIST dataset is much smaller than images used in real-world inference applications, which demand significantly more RAM. To address this, we propose memory optimization techniques, detailed below. We use the MNIST dataset to demonstrate how convolution layer parameters impact RAM usage and show how the proposed techniques allow inference execution within the available RAM. 

\par We have detailed the model used in inferencing task on MNIST data in Algorithm~\ref{alg:CNN_inferencing}. It is important to note that the RAM required for computing $z_j^i, h_j^i, j \in \{1, \dots 4\}$ in a single process increases with the increase in the number of weights at the network layers. This makes executing the inferencing task impractical in restricted resource environments. To address this, we computed $z_j^i, h_j^i, j \in \{1, \dots 4\}$ in eight processes sequentially and saved the output shares from each process to a file. This ensures that each executing process operates on the output shares written by the previous process, preventing any data leakage. We observed both the RAM usage and execution time for each process. Additionally, we noted that computing $h_j^i, j \in \{1, \dots 4\}$ (SecureRelu) consumes less RAM compared to the computation of $z_j^i, j \in \{1, \dots 4\}$ (SecureMul, SecureCNN). The RAM usage shown in Table~\ref{Table:Split Two_Layer_CNN MNIST} denotes the maximum RAM consumed among the different processes in the CNN inference task. The execution time shown in Table~\ref{Table:Split Two_Layer_CNN MNIST} denotes the sum total of the execution times of different processes in the CNN inference task. Next, we present the formulae for dimensions of the convolution layer output and the number of multiplications required to compute each element in the output.

\par The input and output of the convolution layer are three-dimensional matrices. For a given set of input dimensions and convolution layer parameters, the output dimensions are computed using \eqref{eq:cnn_dim}. Here, $i_{ch}$, $i_{row}$, and $i_{col}$ represent the number of channels, rows, and columns in the input, while [$s_{row}, s_{col}$], [$p_{top}, p_{bottom}, p_{right}, p_{left}$], $n_{ker}$, $k_{row}$, and $k_{col}$ represent  strides, padding, number of kernels, number of rows and number of columns in the kernel, respectively. Additionally, $o_{ch}$, $o_{row}$, and $o_{col}$ denote the number of channels, rows, and columns of the convolution layer's output. Note that $o_{ch} = n_{ker}$.
\vspace{-0.2cm}
\begin{align} \label{eq:cnn_dim}
    o_{row}&=   \Bigg \lfloor \frac{i_{row}+p_{top}+p_{bottom}-k_{row}}{s_{row}} \Bigg \rfloor +1 \nonumber \\
    o_{col}&=   \Bigg \lfloor \frac{i_{col}+p_{left}+p_{right}-k_{col}}{s_{col}} \Bigg \rfloor +1 
\vspace{-0.1cm}
\end{align}
\vspace{-0.2cm}
\par We supply MNIST data with dimensions $i_{ch} = 1$, $i_{row} = 28$, and $i_{col} = 28$ as input to CNN1. Using \eqref{eq:cnn_dim}, we compute the output dimensions of CNN1 and CNN2, for the parameters outlined in Table~\ref{Table:Split Two_Layer_CNN MNIST}. The resulting output dimensions for CNN1 are $o_{ch} = 5$, $o_{row} = 13$, and $o_{col} = 13$. For each element in the output of CNN1, the number of multiplications to be performed is $i_{ch} \times k_{row} \times k_{col}$, which is equal to $25$. To compute the output of CNN1, the number of multiplications is given by $21,225$. Following the same procedure, we compute the number multiplications at CNN2 as $8,640$ (see the first row in Table~\ref{Table:Split Two_Layer_CNN MNIST}). Further, the number of multiplications at fully connected neural networks layers, NN1 and NN2, are $ 10,800$ and $1000$, respectively (executed using SecureMul()).

\par In MOTION2NX framework, each multiplication operation entails OT transfers that consumes significant RAM. As illustrated in Table~\ref{Table:Split Two_Layer_CNN MNIST}, the RAM usage increases with an increase in the number of multiplications, posing a challenge to providing a  scalable solution. To address this scalability issue at the convolution layer, we adopt a kernel-by-kernel (vertical split) approach, conducting computations for each kernel individually in a sequential manner. After convolution operation with each kernel, we save the output shares to a file. Upon completing of the convolution operation for all the kernels, we concatenate the output shares obtained kernel-wise sequentially, creating a final output share file. 
\par We have two convolution layers CNN1 and CNN2 with number of kernels equal to $5$ and $3$ respectively. We execute inferring task using different split configurations for the model described in Table~\ref{Table:Two_Layer_CNN_parameters} and tabulate RAM usage and executions times in Table~\ref{Table:Split Two_Layer_CNN MNIST}. The first column depicts the split configuration we used at the CNN layers, while the second column shows the number of splits at the fully connected NN layers. For example, CNNV split = $(5, 1) $ refers to  executing convolution at CNN1 with $5$ kernels, kernel by kernel sequentially and executing the whole convolution operation in one process at CNN2. Further, NN split = $(20, 2)$ means executing matrix multiplication at NN1 and NN2 in $20$ and $2$ splits respectively~\cite{Burra}.
\vspace{-0.2cm}
\begin{remark}
    In any layer, whether it be a convolution or a fully connected neural network layer, having a single split means that all computations related to that layer are computed in a single process.
\end{remark}
\vspace{-0.2cm}
\par  To examine the impact of splits on convolutional layers, we explored various split configurations and calculated the corresponding RAM usage and inference times, as shown in Table \ref{Table:Split Two_Layer_CNN MNIST}. A noteworthy observation was that, for a given number of multiplications, matrix multiplication consumes more RAM than convolution. To explain this, consider the first and second entries for the CNNV Split and NNSplit configurations: $(1, 1), (1, 1)$ and $(1, 1), (20, 2)$, respectively. In the first entry, the maximum number of multiplications occur at CNN1 (=$21,125$) in convolution layers and at NN1 (=$10,800$), with RAM usage of $0.219$ GB. Let’s assume that CNN1 predominantly contributes to the overall RAM usage. In the second entry, the execution is split at NN1 and NN2 (NNSplit $=(20, 2)$) with no split at CNN1 and CNN2. If CNN1 was responsible for the maximum RAM usage, then the RAM usage in the second entry would not have decreased. However, we observe a reduction in RAM usage to $0.092$ GB, indicating that NN1, despite having fewer multiplications ($10,800$) compared to CNN1 ($21,125$), is the primary driver of RAM consumption.
\par In third, forth and fifth entries we use NNSplit $ = (20, 2)$, but changed splits at convolution layers and indicated the split (CNNV) and the number of multiplications that drives the RAM usage in bold. We clearly observe that as the number of multiplications reduced the RAM usage also reduced. 
\vspace{-0.3cm}
\paragraph{Helper Node:} We observe that in secure convolution and secure matrix multiplication, the execution time is significantly impacted by the use of oblivious transfers (OTs) which occur behind the scenes. To address this issue, we use a semi-honest third-party Helper node that eliminates the need for OTs during convolution. For detail discussion on Helper-node algorithm see \cite{Burra}. Observe that the Helper-node algorithm requires an inference time of only $15.2$ seconds, while the RAM requirement is $0.042$~GB (see Table~\ref{Table:Split Two_Layer_CNN MNIST}). Note that, while the Helper node functionality reduces both execution time and RAM usage, it requires an additional server, thereby increasing the server count from two to three.
\vspace{-0.2cm}
\begin{table}
\caption{Ram Usage and Execution Time for Different Split Configurations for MNIST Data Inferring Task}
\centering
\begin{tabular}{|c|c|c|c|c|c|c|c|c|c|} 
        \hline
        \multicolumn{2}{|c|}{split configuration}&\multicolumn{4}{|c|}{No. of multiplications}&\multicolumn{2} {|c|}{LAN}&\multicolumn{2} {|c|}{WAN}\\
       \cline{1-10}
       {}&{}&{} & {} & {} & {} & {RAM} & {time}& {RAM} & {time}\\ 
       {CNNV split}&{NNSplit}&{CNN1} & {CNN2} & {NN1} & {NN2} & {(GB)} & {(sec)}& {(GB)} & {(sec)}\\
        \hline
(1, 1)&(\textbf{1},1)& 21,125 & 8,640 & \textbf{10,800} & 1000 & 0.219 & 19.1 & 0.219 & 20.7 \\ \hline
(\textbf{1}, 1)& (20,2)&\textbf{21,125} & 8,640 & 540 & 500 &0.092 & 31.1 & 0.089 & 42.4 \\ \hline
(5, \textbf{1})& (20,2) &4,225 & \textbf{8,640} & 540 & 500 & 0.057 & 26.3 & 0.058 & 50.2 \\ \hline
(\textbf{1}, 3)&(20,2) &\textbf{21,125}& 2,880 & 540 & 500 & 0.109 & 20.3 & 0.109 & 46.5 \\ \hline
(\textbf{5,} 3)& (20,2)&\textbf{4,225} & 2,880 & 540 & 500 & 0.035 & 43.1 & 0.036 & 54.3\\ \hline

\hline
\hline
\multicolumn{2}{|c}\textbf{Helper node} &21,125 & 8,640 & 10,800 & 1000& 0.042& 14.1 & 0.042  & 15.2 \\
\hline
\end{tabular}

\label{Table:Split Two_Layer_CNN MNIST}
\vspace{-0.3cm}
\end{table}

\subsection{Horizontal Split: }\label{subsec : horiz_split}
\vspace{-0.1cm}
\begin{algorithm}
    \caption{Horizontal Split: Computation of the set of row-start-indices and row-end-indices for all the horizontal splits of the input data}
    \label{alg:Horiz-split}
     \hspace*{\algorithmicindent} \textbf{Input:} Padding values $[p_{top}, p_{bottom}, p_{left},p_{right}]$, stride values $[s_{row},  s_{col}]$, number of rows in each input channel $i_r$, number of rows in each kernel $k_{row}$, number of horizontal splits $n_h$\\
    \hspace*{\algorithmicindent} \textbf{Output:} Set of row-start-indices and row-end-indices for all the horizontal splits of the input data while accounting for padded rows
    \begin{algorithmic}[1]
        \State \emph{Pre-processing}: Update the input data by appending rows and columns (with zero entries) based on the padding values $[p_{top}, p_{bottom}, p_{left},p_{right}]$
        \State Compute the effective number of rows after padding (on updated input)\\
        \quad $D_r = i_r + p_{top} +p_{bottom}$
        \State Compute the number of rows per channel at the output of  the  convolution: \\
        \quad $o_r = \lfloor\frac {D_r - k_{row}}{s_{row}}\rfloor + 1$

        \State Compute the number of output rows for each horizontal split, except the last one: \\
        \quad $ h_r = \lfloor \frac{o_r}{n_h}\rfloor  $
       \State Make necessary initializations, and proceed with the computation:\\
       \quad $ S_r[i] = 0 $ for $i \in \{1,\ldots,n_h\}$ \Comment{row start indices} \\
       \quad $ E_r[i]= 0 $  for $i \in \{1,\ldots,n_h\}$ \Comment{row end indices}\\
       \quad $t = 0$
       \For{$i \in \{1, \dots ,n_h \}$}
            \If{$i == n_h$}
            \State $ h_{r} = \lfloor \frac{o_r}{n_h}\rfloor +  (o_r \mod n_h) $
            \EndIf
            \If{$i == 1$}
            \State $ S_r[i] = 1 $
            \Else 
            \State $ S_r[i] = t-(k_{row}-1)+s_{row}$
            \EndIf
        \State $E_r[i] = S_r[i] + (k_{row} - 1)+ (h_r-1)*s_{row} $
        \State $t = E_r[i] $
      \EndFor 
    \end{algorithmic}
\end{algorithm}

In practical image inference applications e.g, X-ray inference, the image size would of the order of megabytes, and even the kernel-by-kernel execution of the secure inference task may not be feasible under given RAM constraints. To address this, we divide the input data at the convolution layer into smaller chunks. We then perform the convolution sequentially on these smaller input data chunks kernel-by-kernel and append the outputs obtained from each chunk to form the final output. We refer to this task of splitting the input data (row-wise) into smaller chunks as ``horizontal splitting". We explain this in detail in Algorithm\ref{alg:Horiz-split}.

\par The inputs to Algorithm~\ref{alg:Horiz-split} are the number of rows for each input channel at CNN layer, padding values, stride values and number of horizontal splits on the input at CNN layer. The output of Algorithm~\ref{alg:Horiz-split} is the set of row-start-indices and row-end-indices for the horizontal splits of the input data. In step~$1$ we pre-process the input data by appending rows and columns (with zero entries) based on the padding values $[p_{top}, p_{bottom}, p_{left},p_{right}]$. In step~$3$ we compute the total number of rows in the input after padding. In step~$5$ we compute the total number of output rows in one channel at CNN layer output corresponding to the pre-processed input dimensions. In step~$7$ we compute the number of output rows that will be present for the given number of horizontal splits $n_h$. In steps~$9$ and $10$ we initialize the arrays that store start and end row indices of the horizontal splits and in steps $12$-$22$ we compute the indices. 

Once the indices are computed, for each start index and end index, we read the corresponding data from the updated input on which convolution has to be performed. 
We remark that this horizontal splitting algorithm provides a reduction in RAM usage roughly by a factor of $n_h$, where $n_h$ denotes the number of horizontal splits. Further, this splitting algorithm is applicable to general SMPC protocols.
\vspace{-0.2cm}
\section{Conclusion and Future Work}
\vspace{-0.2cm}
We modified and enhanced the MOTION2NX framework to bridge the gap between scalability, memory efficiency and privacy. In particular, we optimized the RAM usage and the execution time via several enhancements to MOTION2NX, while preserving data privacy. These
optimizations enable MNIST dataset inference in just $15$ seconds with only $42$~MB of RAM usage for a four layer CNN. In contrast, the baseline implementation where we executed the image inference task as a single process required $19.7$ seconds of execution time and $800$~MB of RAM. An interesting future work is to use the Helper node for implementing activation functions and solving linear optimization on MOTION2NX.

\vspace{-0.2cm}

\bibliographystyle{splncs04}
\bibliography{cnn}

\end{document}